\documentstyle[12pt]{article}

\newcommand{\nc}{\newcommand}  
\def\frac#1#2{{\textstyle {#1 \over #2}}}

\nc{\beq}{\begin{equation}}   
\nc{\eeq}{\end{equation}}   
\nc{\beqa}{\begin{eqnarray}}   
\nc{\eeqa}{\end{eqnarray}}   
\nc{\lsim}{\begin{array}{c}\,\sim\vspace{-21pt}\\< \end{array}}   
\nc{\gsim}{\begin{array}{c}\sim\vspace{-21pt}\\> \end{array}}   
   
\def\&{and}

\def\nc#1#2#3{           {\it Nuovo Cim.  }{\bf #1}, #2 (19#3)}

\begin{document}   
   
\begin{titlepage}   
\begin{center}   
\vskip .5 in 
\hfill CPTH-RR-034-0302 
\par\hfill Bicocca-FT-02-04 
\par \hfill SHEP-0205
\par
\hfill hep-th/0203203 
 
\vskip .3 in    
{\Large \bf The Gravity Dual of Softly Broken ${\cal N}=1$ \\ \vspace{0.3cm}    
Super Yang-Mills}   
\vskip .3 in   
    
    
\centerline{\bf Nick Evans{$^\flat$},    
Michela Petrini$^\sharp$, Alberto Zaffaroni$^\natural$}    
\bigskip     
\bigskip    
\bigskip     
\centerline{\it $^\flat$Department of Physics, University of Southampton}    
\centerline{\it Southampton SO17 1BJ, U.K.}   
\centerline{\small \tt evans@phys.soton.ac.uk}   
\centerline{$\phantom{and}$}    
\centerline{\it$^\sharp$Centre de Physique Th{\'e}orique, Ecole Polytechnique}   
\centerline{\it F-91128 Palaiseau cedex, France}   
\centerline{\small \tt  michela.petrini@cpht.polytechnique.fr}   
\centerline{$\phantom{and}$}   
\centerline{\it $^\natural$ Dipartimento di Fisica, Universit{\'a} di Milano-Bicocca}   
\centerline{Piazza della Scienza 3, I-20126, Milano, Italy}   
\centerline{\small \tt alberto.zaffaroni@mib.infn.it}   
   
\end{center}   
\vskip .5 in   
   
\begin{abstract}   
   
\noindent   
   
Starting from the Maldacena-Nunez supergravity dual of ${\cal N}=1$   
super Yang-Mills   
theory we study the inclusion of a supersymmetry breaking gaugino mass term.   
We consider a class of 
non supersymmetric deformations of the MN solutions which have been recently 
proposed in the literature.  
We show that they can  be 
interpreted as corresponding to the 
inclusion of both a mass and a condensate.   
We calculate the vacuum energy of the supergravity solutions showing   
that the $N_c$-fold vacuum degeneracy of the  ${\cal N}=1$ theory is lifted   
by the inclusion of a mass term.   
\end{abstract}   
   
\end{titlepage}   
   
\input epsf   
\newwrite\ffile\global\newcount\figno \global\figno=1   
\def\writedefs{\immediate\openout\lfile=labeldefs.tmp \def\writedef##1{%
\immediate\write\lfile{\string\def\string##1\rightbracket}}}   
\def\writestoppt{}\def\writedef#1{}   
\def\figin{\epsfcheck\figin}\def\figins{\epsfcheck\figins}   
\def\epsfcheck{\ifx\epsfbox\UnDeFiNeD   
\message{(NO epsf.tex, FIGURES WILL BE IGNORED)}   
   
\gdef\figin##1{\vskip2in}\gdef\figins##1{\hskip.5in}
\else\message{(FIGURES WILL BE INCLUDED)}%
\gdef\figin##1{##1}\gdef\figins##1{##1}\fi}   
\def\figinsert{}   
\def\ifig#1#2#3{\xdef#1{fig.~\the\figno}   
\writedef{#1\leftbracket fig.\noexpand~\the\figno}%
   
\figinsert\figin{\centerline{#3}}\medskip\centerline{\vbox{\baselineskip12pt   
\advance\hsize by -1truein\center\footnotesize{  Fig.~\the\figno.} #2}}   
\bigskip\endinsert\global\advance\figno by1}   
\def\footnotefont{}\def\endinsert{}   
   
\renewcommand{\thepage}{\arabic{page}}   
\setcounter{page}{1}   
   
\section{Introduction}   
   
The AdS/CFT correspondence \cite{ads}    
provides the prototype example of a duality    
between a gauge theory (${\cal N} = 4$ Super Yang-Mills (SYM) in four   
dimensions) and a string theory (type IIB on the background $AdS_5    
\times S^5$). There has been great interest in trying to extend   
this duality to other less supersymmetric gauge theories. Much attention has   
focused on theories with less, but non zero, supersymmetry, where    
gauge theory is more clearly understood. Supergravity backgrounds have been   
found both by the deformation \cite{deform}   
of the ${\cal N} = 4$ theory by relevant    
perturbations and from consideration of the near horizon limit of more   
complicated D-brane constructions \cite{construct}.    
For example, the ${\cal N}=2^*$ theory    
has been obtained by including masses that break ${\cal N} = 4$    
to ${\cal N} = 2$ in the infrared \cite{N2*},    
while the pure ${\cal N} = 2$ theory has been realized from   
D5 branes wrapped on a two cycle \cite{N2} or from branes at an orbifold 
singularity \cite{N2f}.     
The ${\cal N} = 1^*$ theory has been   
studied both as a deformation \cite{N1} and by using fractional branes   
\cite{ks}. In this paper we shall focus on the dual of ${\cal N} = 1$ SYM    
provided by Maldacena and Nunez resulting from D5 branes wrapped on a two     
cycle \cite{MN}, based on the solution in \cite{CV}.
 In each of these cases it has been understood how   
the gravity    
dual reproduces the perturbative running of the gauge theory though     
non-perturbative aspects of the theory are more hit and miss depending  
on what phenomena survive the large $N_c$ limit. For example    
the Maldacena-Nunez ${\cal N} = 1$ solution includes a gaugino condensate     
whilst in the ${\cal N} = 2$ theory instanton effects are squeezed into     
an enhan\c con singularity  \cite{enhance} of the supergravity background.    
    
Dualities with string theories appear to be a robust phenomena in   
gauge theory    
and it is therefore interesting to now try to extend our understanding    
beyond supersymmetric theories. One might hope eventually to have a description    
of strongly coupled Yang Mills theory or even QCD. A few steps have already    
been taken in this direction, including     
studying non-supersymmetric deformations    
of the  ${\cal N} = 4$ theory with 5-dimensional supergravity     
\cite{listdef}, type 0 theories \cite{typeO} and finite temperature    
theories \cite{finiteT}\footnote{See \cite{verlinde} for the study of a metastable $N=0$ solution.}. In the interesting case of zero temperature,    
all these solutions are plagued by singularities.     
More recently the authors of \cite{aharony} studied the inclusion of   
supersymmetry breaking     
scalar operators in the Maldacena-Nunez ${\cal N} = 1$ theory, which    
results in a regular background. In this paper, we wish to    
study the inclusion of  a gaugino mass term rather than scalar     
operators. Again a set of IR regular solutions can be found.    
The gaugino mass term has the interesting property of breaking     
the $U(1)_R$ symmetry of the ${\cal N} = 1$ theory. The result of this, as    
we show by calculating the vacuum energy of the appropriate     
supergravity backgrounds, is to lift the $N_c$-fold vacuum degeneracy of     
$SU(N_c)$ ${\cal N} = 1$ SYM. Much of the computational technology we shall use    
has already been studied in \cite{gubser} where non BPS versions of the    
Maldacena-Nunez solution were exhaustively presented. The purpose 
of \cite{gubser} was different from ours, the authors 
being interested in finite energy excitations of the MN solution, with zero 
or finite temperature. We show that a class of solutions in \cite{gubser} 
can be interpreted as a gravity dual of softly broken ${\cal N}=1$ SYM.  
In order to be self-contained, we will review most of the computation 
in \cite{gubser}. 
We will    
then 
make contact between these gravity solutions and the physics of the dual     
gauge theory.    
     
Let us review the field theory expectations when a gaugino mass     
is introduced into ${\cal N} = 1$ $SU(N_c)$ SYM.     
The ${\cal N} = 1$ theory is known to have a mass gap    
due to the formation of a gaugino condensate and the vacuum is therefore     
described by the holomorphic superpotential    
\beq     
W = \Lambda^3  e^{i n 2 \pi / N_c}, \hspace{1cm} n=0..N_c-1,     
\label{n1w}   
\eeq    
where $\Lambda$ may be written in terms of the bare coupling $\tau = {\theta     
\over 2 \pi} + i {4 \pi \over g^2}$ at some UV scale  as    
\beq    
\Lambda = \Lambda_{UV} e^{2 \pi i \tau / 3N_c}.    
\eeq    
    
The $N_c$-fold degeneracy of the vacuum corresponding to the $N_c$     
choices of phase in (\ref{n1w}) is related to the anomalous     
breaking of the     
U(1)$_R$ symmetry of the theory to $Z_{2N_c}$ by instanton effects    
(the $Z_2$ symmetry on the gaugino    
$\lambda \rightarrow - \lambda$ is left unbroken by the bifermion condensate).    
    
Soft breaking terms may be introduced into supersymmetric theories by    
allowing the parameters of the theory to have non-zero $F$-components    
\cite{soft}. If $\tau$ has a non-zero $F$-component, $f_\tau$,     
then in the bare lagrangian    
a gaugino mass is introduced. If the mass is small relative to    
$\Lambda$ the supersymmetry breaking term will act as a perturbation to the     
stable $N_c$ vacua and the resulting theory will still be described by      
(\ref{n1w}). Accounting for $f_\tau$ it can be seen that the vacuum energy    
is no longer zero but, at leading order in the mass, it is    
 given by (for $\theta=0$ and $f_\tau$ real)    
\cite{softym}    
\beq \label{pot}    
\Delta V = - 32 \pi^2 m_\lambda \Lambda^3 \cos \left[ { 2 \pi n \over N}     
\right].    
\eeq    
The plane of the vacua is tilted and there is a single unique vacuum ($n=0$).    
    
This is the property of the softly broken theory we uncover in the     
supergravity dual below. In the next section we will review the   
Maldacena-Nunez    
solution and identify the field corresponding to the gaugino condensate.  
In section 3 we show how a more general solution of the second order   
supergravity    
equations allows the inclusion of a mass term for the gaugino.   
Section 4 describes the     
determination of the vacuum energy of the spacetimes describing   
the perturbed vacua    
of the ${\cal N} = 1$ theory and reproduction of  
 the field theory result eq. (\ref{pot}).

\section{Gravity Dual of ${\cal N} = 1$ SYM}    
    
Consider a wrapped five brane with     
world-volume $R^4\times S^2$. This general setting can be    
easily adapted to describe both ${\cal N} = 2$, ${\cal N} = 1$    
and ${\cal N} = 0$ theories.    
   
In the case of $N_c$ flat NS5 branes the world-volume theory is a    
little string theory which reduces in the IR to  ${\cal N} = 1$    
six-dimensional SYM theory.  
The theory contains 4 scalars transforming in the 
$4$ of the     
$SO(4)_R$ R-symmetry group, and two symplectic  
Majorana fermions transforming in the $(4,2)+(4^\prime,2^\prime)$  
of $SO(5,1)\times SO(4)$.      
Wrapping the brane on an $S^2$ one    
obtains 4-dimensional gauge theories with coupling inversely proportional     
to the volume of the $S^2$\footnote{The presence of the  $S^2$ naturally   
implies the existence of Kaluza Klein modes in the theory. Since the theory is   
at large $g^2_{YM} N_c$ these massive modes can not be considered decoupled   
and in this  sense the dual gauge theories are not the pure 4 dimensional   
theories one  might have hoped for. Nevertheless we hope they lie in the same    
universality  class.}.   
Since there are no covariantly constant spinors on $S^2$,    
supersymmetry is generally broken by the compactification.     
In order to preserve some supersymmetry the theory has to be twisted, namely    
the spin connection on $S^2$   
has to be identified with a background $U(1)_R$ field   
in the $SO(4)_R$  R-symmetry group \cite{vafa,MN}.    
This can be easily seen      
from the supersymmetry variation of a fermion    
\begin{equation}    
\delta \Psi\sim D_{\mu}\epsilon=(\partial_{\mu}+\omega_{\mu}^{\nu\rho}\gamma^{\nu\rho}    
-A_{\mu})\epsilon.   
\label{twist0}    
\end{equation}    
The choice of the $U(1)$ in $SO(4)_R$ determine the amount of the   
supersymmetry left. For ${\cal N}=1$ supersymmetry the relevant twisting    
$U(1)_R$ is the abelian subgroup of $SU(2)_R$    
in the decomposition $SO(4)\rightarrow SU(2)_R\times SU(2)_L$    
\footnote{Solutions with ${\cal N}=2$ supersymmetry have been discussed in    
\cite{N2}. In that case, the $U(1)$ field   
is a combination of the two abelian subgroups $U(1)_L+U(1)_R$.}.  
 
As is standard in the AdS/CFT correspondence,   
the $SO(4)$ gauge fields correspond  
to the isometries of the 3-sphere and are dual to the R-symmetry group.  
In the 7 dimensional field theory we are including a source term with  
the symmetry properties of a U(1)$_R$ gauge field.  
  
The massless    
four dimensional fields in the ${\cal N}=1$ theory are the gluons    
and the gluino $\lambda$.    
The latter is the only component of the $S^2$ reduction    
of the five-brane fermionic field $\Psi$ satisfying the twist    
condition: $(\omega_{\mu}^{\nu\rho}\gamma^{\nu\rho} -A_{\mu})\epsilon=0$.   
All other scalars and fermions from the reduction of the six dimensional    
theory acquire mass due to the twist.    
    
The dual supergravity background has been constructed, as usual,  
 by first reducing to a  
lower dimensional gauge supergravity and then by  
lifting the solution to 10 dimensions.  
In this case, the relevant theory is 7-dimensional $SO(4)$   
gauged supergravity, which corresponds to the truncation   
of the type I sector of type IIB on the 3-sphere transverse to the NS5 brane.  
This is a consistent choice   
since the NS5 branes only couple to the NS sector of type IIB  
SUGRA \cite{MN}. in fact the 
solution of \cite{MN} is the lift to 7 dimension of a non-abelian monopole solution 
of 4 dimensional $SU(2)$ gauge supergravity found in \cite{CV}.
In a 7 dimensional string frame the solution describing wrapped NS5 branes consists of the    
the metric  
\beq    
ds^2_{7} =  dx_4^2 + {N  \over 4} \left[ dr^2 +     
e^{2g}(d \theta^2    
+ \sin^2 \theta d \phi^2) \right],    
\eeq    
the $SU(2)_R $ gauge fields    
\beq   \label{gauge} 
A = {1 \over 2} \left[ \sigma^3 \cos \theta d \phi +   
{a\over 2} (\sigma^1 +i\sigma^2) (d \theta  -i\sin \theta d \phi) + \mbox{c.c}  
\right],    
\eeq    
and the dilaton   
\beq    
e^{ 2 \Phi_{NS}-2\Phi_{NS0}} = { e^{g}\over \sinh r}.   
\eeq    
The functions    
\beqa    
a &=& { r \over \sinh r} e^{i \chi}, \nonumber\\   
e^{2 g} &=& 2r \coth r - |a|^2-1  
\eeqa    
have been obtained from the supersymmetry variations in the 7-dimensional   
theory \cite{MN}.  Here $a$ is a complex field whilst $g$ and $\Phi_{NS}$ are  
real.  The solution has non vanishing non abelian gauge fields, which go   
to zero in the UV and reduce to a pure gauge in the IR. Their inclusion  
is required in order to have a regular solution and it 
breaks the $U(1)_R$ symmetry of the theory to $Z_2$ in the IR.    
To describe a decoupled four dimensional theory we need to go to a D5 description \cite{MN}. The supergravity 
solution for a wrapped D5 is obtained by S-duality from the NS5 solution.  
In the 
next sections we will always consider wrapped D5's. 
 
The supergravity fields entering the Maldacena-Nunez solution    
are dual to four dimensional composite operators.     
In particular the field $a$, which is necessary  
in order to have a regular solution, is dual to the gluino    
condensate.    
This follows from the gauge fields in (\ref{gauge}) and 
the $S^2$ reduction of the     
six dimensional lagrangian term  \cite{boxing}    
\beq    
\bar \Psi A_\mu \gamma^\mu \Psi\qquad\rightarrow    
\qquad a \bar\lambda\lambda.    
\label{glu}    
\eeq       
  
As usual in the AdS/CFT correspondence,    
the UV gauge theory is determined by the large $r$ behaviour of the solution.    
At large $r$ we find    
\beq    
a(r)\sim K r e^{-r},   
\label{condensate}    
\eeq    
where $K=2 e^{i \chi}$ should be interpreted as (the complex conjugate of) 
the condensate \cite{boxing}.     
Note that in the equation above we consider the full complex $a$ field    
since it allows us to describe the gaugino condensate including   
its phase, $\chi$. In this formula, the condensate has a free phase.  
The anomaly restricts     
$\chi$ to discrete values $2 \pi n/N_c$ corresponding to the $N_c$ vacua.  
The anomaly has been identified in the behaviour    
of the antisymmetric field $B_{NS}$     
\cite{klebwit} and in the contribution of world-sheet instantons    
\cite{MN} in the 10d lift of the solution. As shown in \cite{MN},  
only these discrete values of $\chi$ give rise to fully consistent    
10 dimensional  
solutions. By considering the IR part of the metric, one can  
see that each of them only preserves a $Z_2$ symmetry, consistently  
with QFT expectations.   
  

\section{A Soft Breaking Mass}    
    
To introduce a mass term into the ${\cal N} =1$ solution we note that    
a gaugino mass and a gaugino condensate share the same symmetry properties
(up to conjugation)     
and should therefore both be described by a more general solution for    
$a$.    
Since a mass term for the gaugino breaks supersymmetry, on the supergravity   
side we can no longer look at the variations of the fermions, but we need    
to solve the second order supergravity equations of motion. Fortunately,   
much of the work has been done, for a different purpose, in    
\cite{gubser}. In particular, what we need is the 
1-dimensional    
effective Lagrangian describing the fields we are interested in \cite{gubser} 
\beq    
{\cal L}=e^{2s}\left (s^{\prime 2}-{e^{-2g}\over 2} |a^{\prime}|^{2}-{g^{\prime 2}\over 2}-{1\over 4}[e^{-2g}(|a|^2-1)^2-2e^{-2g}-1]\right ).    
\label{eff}    
\eeq     
Here $s=\Phi+g$ and $\Phi=-\Phi_{NS}$ is the dilaton in the D5 description.  
The second order equations of motion are then given by    
\beq   
\begin{array}{l}    
\left[ \  e^{2s-2g} a^\prime \ \right]^\prime = e^{2s-4g} (|a|^2-1) \ a, \nonumber\\  
\\   
\left[ \   e^{2s} (s^\prime ) \ \right]^\prime = {1 \over 2} e^{2s}     
\left[-  e^{-4 g} (|a|^2 -1)^2 +2 e^{-2 g} + 1\right],  \nonumber\\   
\\  
\left[ \ e^{2s} g^\prime \ \right]^\prime =  e^{2s} \left[ -e^{-4 g}     
(|a|^2 -1)^2 + e^{- 2g}\right] - e^{2s-2g} \ |a^{\prime}|^{2}.     
\end{array}   
\eeq    
These equations admit the supersymmetric solution, which corresponds to the   
Maldacena-Nunez background.    
The most general solution to the second order equations of motion for $a$,    
which we present in full below,      
admit at large $r$, a     
non-normalizable asymptotic solution $a_1 \sim 1/\sqrt{r}$ and     
a normalizable one $a_2 \sim re^{-r}$. As we have seen above a  background    
where only the normalizable solution is turned on is associated with    
a vacuum of the field theory with a VEV for the corresponding operator.    
The non-normalizable solution $a_2$ changes the UV behaviour of the     
solution and it is therefore associated with a deformation of the theory    
where the gaugino has a mass \footnote{We might expect      
the relative scaling dimension of the two sources to be apparent from    
the $r$ dependence of the solution in the UV (for example, naively,     
one scaling as $r$ the other as $r^3$) but in this case there does     
not seem to be a straightforward interpretation.    
}.   
     
In general, therefore,  the solutions     
of the second order equations,    
with two complex free parameters associated with $a$,    
will describe both the gaugino condensate and a mass term for the gaugino.    
These solutions will generically break supersymmetry.    
The full asymptotic behaviour for large $r$ reads    
\begin{eqnarray}    
a &=& {Y\over \sqrt{r}} (1 + {1-|Y|^2/2\over r}+...) + C r e^{-r}(1+{\gamma\over r}+...),\nonumber\\    
g &=& {1 \over 2} \log 2 r -{|Y|^2\over 2r^2}(1+...) +   
P \sqrt{r} e^{-r}(1 + { \alpha \over r}+...),\\    
\Phi &=& \Phi_0 + r/2 - \log{r}/4 + {5|Y|^2\over 16r^2}(1+...) - P' \sqrt{r} e^{-r}  ( 1 + {\beta \over r}+...),  \nonumber  
\label{asymp}    
\end{eqnarray}    
where dots stand for corrections in $1/r$.    
$\Phi_0$ is a free parameter determining the dilaton (coupling)     
at a given $r$ (scale).    
We interpret the complex parameters 
$Y$ (the non-normalizable solution) as the mass deformation,  
and    
$C$ (the normalizable solution) as the complex conjugate of the  condensate.  The functions $P$,     
$P'$, $ \alpha$, $\beta$ and $\gamma$    
are determined by the equations of motion and we find    
\beq    
P = P' = k Re(\bar{C} Y), \hspace{0.7cm} \alpha = 2 + {1 \over 2 k},  
\hspace{0.7cm}    
\beta = 1 + {1 \over 2 k}, \hspace{0.7cm} \gamma={1+(4k+3)|Y|^2\over 2},    
\label{parameters}  
\eeq    
where $k$ is a free parameter \footnote{Note that our solution differs a 
little from that in \cite{gubser}. They present as the solution the limit of 
our equations where $k \rightarrow \infty$ with $k Re( \bar{C}Y) \rightarrow {\rm constant}$. The parameter $P$ then becomes free.  In fact the result that $P$ is proportional to $Re(\bar{C}Y)$ is crucial to  our analysis of the vacuum energy below.}.   
Finally    
we would expect two other free parameters in the solution of these equations    
which are encoded in the freedom to shift and scale the $r$ coordinate    
\beq  
r \rightarrow \mu (r + r_*).  
\eeq    
Clearly there are many more free parameters than we expect in the     
field theory which is uniquely determined by the UV value of the mass    
and the coupling. However from the analysis of the IR solutions \cite{gubser}  
we see  that there is a two parameter family of {\it regular} solutions    
given by  
\beqa   
a &=& e^{i \chi'} (1 - b r^2 + ...), \nonumber\\    
e^g &=&r - ( b^2 + {1 \over 36}) r^3 + ...,\\   
\Phi& =& \phi(0) + (b^2 + {1 \over 12}) r^2 + ....\nonumber    
\eeqa    
Restricting to these solutions, the eight real parameters in the UV are reduced    
to three. $\phi(0)$ matches to $g^2_{YM}$, and $b$ to the gaugino mass term.
The regular IR solution has only a single phase in the field $a$ whilst
the UV solution has two, one on $Y$ and one on $C$. Regularity in the IR therefore
forces the phase of the condensate and the conjugate of the mass term to be equal.
This is precisely the condition for the minimum of the field theory potential
in (\ref{pot}) and is our first hint that the gravity solution correctly
encodes the field theory.

The full solutions can be found     
by numerically integrating the IR solutions to the UV and solving for     
the UV parameters as a function of $b$ and $\phi(0)$.  
In the range $b\in [0,1/6]$, the solutions have a regular behaviour \cite{gubser} and we can interpret them as mass deformations of the MN solution. 
At the supersymmetric point in the IR $b=1/6$ whilst in the  
UV $\mu=1, r_*=-1/2, \ C=2/\sqrt{e}$ and $Y=0$ ($\Phi_0$ or $\phi(0)$   
is a  free choice).   
In fact to determine the vacuum energy of these   
configurations we shall only need the UV asymptotic forms of the solutions.   
There is a subtlety though; we will need to know the value of the parameter  
$k$ when we break supersymmetry. At the supersymmetric point $k$ is  
undetermined because $Y=0$, however, its value can be found as the  
limiting value of $k$ as $b \rightarrow 1/6$.  This requires the  
numerical integration procedure described above.   
The numerical procedure necessarily runs into trouble  
at $b=1/6$ but the limiting value can be read off to be approximately  
-390. All we will need is that $k$ is a constant plus ${\cal O} (b) \sim  
{\cal O}(m_\lambda)$.  Note that $k$ is a $U(1)_R$ invariant
quantity and hence has the same value independent of the phases
on $Y$ and $C$.
%
%

Finally we recall that these     
solutions can be lifted to ten dimensions as for the Maldacena-Nunez  
solution giving    
\beqa    
ds^2_{10 str} &=& e^{\Phi}\left[ dt^2 + dx^2 + {1 \over 4} dr^2 + {1 \over 4}     
e^{2g}(d \theta^2    
+ \sin^2 \theta d \phi^2) + 1/4 \sum_a(w^a - A^a)^2 \right], \nonumber\\   
e^{ 2 \Phi-2\Phi_0} &=& { e^{-g} \sinh r},\\ 
C_{2}^{RR}& =& N \left[ {1 \over 4} (w^1-A^1) \wedge (w^2 - A^2) \wedge     
(w^3 - A^3) - 1/4 \sum_a F^a \wedge (w^a-A^a) \right], \nonumber 
\eeqa    
where $w_i$ is a set of 1-forms describing a three-sphere.    
It is interesting to notice that the deep IR    
form of the metric is exactly the same for all solutions, supersymmetric    
and not. We interpret this as meaning that the theories all possess a  
mass gap    
and the deep IR is therefore the same for all these theories.    
The study of the full solution would reveal the differences in the spectrum    
and the dynamics due to the inclusion of a mass term.

\section{Vacuum Energy}

We have seen that the supergravity equation of motions   
give rise to a set of solutions in the UV differing 
only in the phase of the     
gaugino in the condensate, $C$.  
We will therefore be interested in  
comparing solutions    
of the second order equations with fixed gaugino mass, $Y$, and varying  
phase    
condensate $C$. In    
this section we will compute the relative vacuum energy of these  
space-times     
to determine the true vacuum. As we will see there are contributions to
the vacuum energy both from the UV and the IR of the solution.
The UV term reproduces the expected form in the field theory. For the regular
solution, which we identified above with the true vacuum,
the IR contribution vanishes and we can therefore reproduce
the field theory true vacuum energy's dependence on the breaking mass term.
For the metastable vacua the IR solutions are not regular and the IR 
contribution is less clear. Since the UV contribution matches the field 
theory expectation at leading order in the mass term we expect that the IR 
contribution is subleading for small mass.

    
The value of the Euclidean action for this family of solutions   
can be found in \cite{gubser} and we briefly review the computation.  
Since the 7-dimensional solution has a non  
trivial dependence only on the radial and the 2-sphere coordinates, in the  
following computation we can neglect the contribution of the 3 flat spatial    
directions (they will only provide a divergent multiplicative constant factor).
We then reduce ourselves to the following     
Euclidean 4-dimensional action in the Einstein frame  
\beq    
I = {1 \over 4 \pi} \int_{\cal M} d^4y \sqrt{g} \left( - {1 \over 4} R     
+ {1 \over 2} \partial_\mu \Phi \partial^\mu \Phi + {1 \over 8} e^{2 \Phi}    
F^a_{\mu \nu} F^{a ~ \mu \nu} - {1 \over 4} e^{-2 \Phi} \right)    
- { 1 \over 8 \pi} \int_{\partial \cal M}  d^3y \sqrt{h}K.      
\eeq    
The surface term in the expression above is the contribution from   
the extrinsic curvature  
$ K = \sqrt{g^{rr}} h^{a b} \partial_r h_{a b},   
= e^{-2g-4 \Phi} {\partial \over \partial r} (e^{2g} e^{3 \Phi}),    
$  
where $h_{a b}$ is the 3-dimensional surface metric,     
$h = e^{2 \Phi}(d\tau^2 + e^{2g}(d\theta^2+ \sin^2\theta  
d\phi^2))$. Explicitly the boundary integral reads  
\beq  
I_{bd}=  - {1 \over 2} \lim_{r\to\infty}  e^{- \Phi} \partial_r \ ( e^{2g +3 \Phi}).  
\eeq    
Using the equations of motion the volume term reduces to a surface integral   
since for the regular solutions the $r \rightarrow 0$ 
limit of the integration  
vanishes. The analysis is thus appropriate only to the 
regular non-supersymmetric
solution which as we described above aligns the phases of the mass 
and the condensate 
in the UV. In fact the result we find below reproduces the field theory result
even for the metastable vacua, so most probably, at leading order in $Y$, the 
integrand vanishes in the $r \rightarrow 0$ limit for all the solutions.
Since we can only connect the UV and IR solutions numerically we have not been able 
to directly prove this though.
\beq    
I_{vol} = {1 \over 8 \pi} \int d^4x \  \partial_\mu ( \sqrt{g} g^{\mu \nu}  
\partial_\mu \Phi)    
= \lim_{r\to\infty}{1 \over 2} e^{2g} e^{2 \Phi} \partial_r \Phi.     
\eeq    
The action is therefore given by    
\beq    
I \sim \lim_{r\to\infty} \partial_r (e^{2g} e^{2 \Phi}).    
\eeq    
Notice that it does not explicitly depend on the field $a$. Let us  
use this result to compute the vacuum energy in the softly broken  
theory.      
Substituting in the solution (\ref{asymp}) we find for the leading term    
\beq    
I \sim \lim_{r\to\infty} \partial_r\left     
(  2 \sqrt{r} e^{(r+ 2 \Phi_0)}\right ).     
\eeq      
This is a divergent piece common to all the solutions. This    
piece will cancel when we compare the energies of any two solutions.     
When we make a comparison between two space-times we must be careful    
to make sure that the metric on the boundary is the same for each of the    
two space-times.   
 We shall use the MN zero    
gaugino mass spacetime as our reference geometry. For that metric, using    
the freedom to shift $r$ and $\Phi_0$ we find    
\beq    
I_{BPS} \sim \lim_{r\to\infty}\partial_r \left (  2 {e^{(r+r_*+ 2 \Phi_0 + 2 \Phi_*)}     
\over \sqrt{r+ r_*}} \right ).    
\eeq    
    
In particular, as in \cite{gubser} we must choose the constants $r_*$ and $\Phi_*$ so the    
dilaton and the $S^2$ metric are the same in both cases. In the field theory    
this corresponds to equating the gauge coupling at the UV scale.    
We note that the coefficient of the $S^2$ metric is    
\beq    
e^{2 g+ 2 \Phi} \simeq  2 \sqrt{r} e^{(r + 2 \Phi_0)}(1+...)  
+ 2 e^{2 \Phi_0}    
(2\alpha P - 2 \beta P)(1+...),     
\label{S2}    
\eeq    
where dots stand for polynomials terms in $1/r$. We also note that the    
polynomial corrections to the leading $\sqrt{r}e^{r}$ behaviour only    
depend on $Y$ and not on $C$.

We first compute the vacuum energy at linear order in the mass. Keeping    
only linear terms in $Y$, the matching  gives     
\beq \sqrt{r + r_*} e^{(r + r_* + 2 \Phi_0 + 2 \Phi_*)} =    
\sqrt{r} e^{(r + 2 \Phi_0)}  + e^{2 \Phi_0} (2\alpha P - 2 \beta P),    
\eeq    
\beq     
\sqrt{2(r+r_*)}=\sqrt{2r}+\sqrt{2}P re^{-r}(1+...).    
\eeq    
    
The first equation can be used to fix $\Phi_*$. The second one then fixes    
$r_*=2Pr^{3/2}e^{-r}+...$.    
The energy difference is therefore \footnote{The formula   
$I=2e^{2\phi_0}P$ was obtained in \cite{gubser}.   
The authors of \cite{gubser}  
were, however, interested in solutions with $Y=0$, interpreted as  
non-supersymmetric finite energy excitations of the MN solution, where  
$P$ becomes an extra parameter. In our case, $P$ is fixed in term  
of $Y$ by equation (\ref{parameters})}  
\beq    
\Delta I = e^{ 2 \Phi_0} 2k Re(\bar{C} Y).\label{result}    
\eeq    
    
In the supersymmetric limit where $Y$,  the 
gaugino mass, is zero the     
solutions with different phases on the (complex conjugate of the)
condensate $C$ are degenerate.    
When a gaugino mass is introduced the energy of the vacua to leading  
order in $Y$ (or $m_\lambda$) is given by    
\beq   
E \sim Re ( \bar{C} Y) \sim Re ( m_\lambda \Lambda^3),  
\eeq    
reproducing the field theory result (\ref{pot}).    
Note it was crucial here that $k=$ constant $+ {\cal O}(m_\lambda)$  
as we showed numerically above.  
    
We can repeat the above calculation to higher orders in $Y$ using the    
asymptotics (\ref{asymp},\ref{S2}). The result is     
that divergent terms of the form $\sim |Y|^2e^r/r^{5/2}+...$ appear in the vacuum    
energy. All these terms depend on the mass $Y$ but not on the condensate $C$    
and reflect the fact that the vacuum energy in the softly broken theory    
is infinite (the leading vacuum graph is a fermion loop with two mass    
insertions). They cancel when computing difference in the energy    
for different vacua. The final result for the vacuum energy 
is still given by eq. (\ref{result}).    
As mentioned before, $Y$, $C$ and $k$ are complicated functions of    
the mass parameter, which can be found by matching the UV and IR     
behaviours of the metric. Formula (\ref{result}) therefore encodes     
all higher order corrections in the mass parameter.

\section{Conclusions}    
    
We have studied softly broken ${\cal N}=1$ theories by deforming    
the Maldacena-Nunez solution. We have computed    
the vacuum energy and verified that the ${\cal N}=1$ degeneracy of vacua    
is lifted according to expectations.   
    
Information about condensates and vacuum energy are encoded in the    
subleading UV behaviour of the solution, once parameters and asymptotics are     
fixed by boundary conditions  and regularity in the IR.    
It would be interesting to study other features of the softly    
broken theory  encoded in the full solution or in the IR behaviour,  
for example, to compute the glue-ball spectrum in    
the Maldacena-Nunez solution and in its deformations. Another interesting    
quantity is the ratio of $k$-strings in this model. It was noticed in     
\cite{klebstring} that in the MN solution the ratio of    
tensions follows the sine formula  
    
\beq    
{T_k\over T_{k^\prime}}={\sin k \pi /N\over \sin k^\prime \pi /N}    
\label{sine} \eeq    
found in ${\cal N}=2$ SYM \cite{ds},   
MQCD \cite{strass} and somewhat supported    
by  recent    
lattice computations \cite{pisani}. It appears that, since the string    
tension is   
fixed by the IR behaviour, the string ratio in the    
softly broken theory is the same as in the Maldacena-Nunez solution,    
that is it follows formula (\ref{sine}). 
The sine formula, or mild modifications  
of it,    
are quite commonly realized in stringy inspired models of YM,    
even if it is known that QFT provides some counterexamples to the    
universality of such a formula    
\cite{counterexamples}.    
    
Finally, we should discuss the issue of stability.  
The solutions we considered   
could be unstable, since  supersymmetry is   
not protecting them anymore. However, since the   
${\cal N}=1$ gauge theory has a mass gap, and the MN solution is expected to   
have a discrete tower of normalizable fluctuations, we could expect that,   
at least for small deformations, stability is preserved. A more detailed   
analysis is nevertheless necessary to determine the absence of tachyons  
in the background.   
    
\vskip .1in    
    
\noindent \textbf{Acknowledgments}\vskip .1in \noindent    
    
We would like to thank F. Bigazzi and A. Cotrone for useful discussions. 
AZ is  partially supported by INFN, MURST     
and the European Commission     
wherein he is associated to the University of Padova.   
NE is supported by a PPARC Advanced Fellowship.   
M.P. is  partially supported by  the European Commission  TMR program 
HPRN-CT-2000-00148.

\end{document}